\documentclass[preprint,twocolumn,3p,review]{elsarticle}
\usepackage{float}
\usepackage{amssymb}
\usepackage{amsmath}
\usepackage{amsthm}
\usepackage{graphicx}
\usepackage{array} 
\usepackage[table,xcdraw]{xcolor}
\usepackage[utf8]{inputenc} 
\usepackage[T1]{fontenc}
\usepackage{lipsum}
\usepackage{booktabs,caption}
\usepackage{multirow}
\usepackage[flushleft]{threeparttable} 

\usepackage{lineno,hyperref}
\modulolinenumbers[1]

\usepackage{tabularx}
\usepackage{longtable}
\usepackage{multicol}
\usepackage{multirow}
\usepackage{comment}

\usepackage{graphicx,rotating,color}
\usepackage{epsfig}
\usepackage{geometry}
\usepackage{amsmath}
\usepackage{natbib}
\usepackage{amssymb}
\usepackage{widetext}
\usepackage{cuted}
\usepackage{multicol}
\usepackage{wrapfig}

\newcommand{\ba}{\begin{eqnarray}}
\newcommand{\ea}{\end{eqnarray}}

\journal{Physics Letters B}

\begin{document}
\begin{frontmatter}
\title{Signatures of $\alpha$-clustering in $^{12}$C and $^{13}$C}

\author[label1]{A. H. Santana Vald\'es}
\author[label1]{R. Bijker}

\address[label1]{Instituto de Ciencias Nucleares, Universidad Nacional Aut\'onoma de M\'exico, Apartado Postal 70-543, 04510 Cd. de M\'exico, M\'exico}

\begin{abstract}
We study the cluster structure of $^{12}$C and $^{13}$C in the framework of the cluster shell model. Simple relations are derived for ratios of longitudinal form factors as well as transition probabilities. It is shown that the available experimental data for $^{12}$C and $^{13}$C can be well described by a triangular structure with ${\cal D}_{3h}$ and ${\cal D}'_{3h}$ symmetry, respectively. 
\end{abstract}


\end{frontmatter}

\section{Introduction}\label{ch1_intro}
Recent measurements of new rotational excitations of both the ground state \cite{Fre07,Kirsebom,marin} and the Hoyle state \cite{Free11,Itoh,Freer12,Gai} in $^{12}$C have led to a renewed interest in the cluster structure of light nuclei, in particular of $\alpha$-conjugate nuclei as clusters of $k$ $\alpha$-particles. Early work on $\alpha$-cluster models goes back to studies by Wheeler \cite{wheeler}, and Hafstad and Teller \cite{Teller}, followed by later work by Brink \cite{Brink1,Brink2} and Robson \cite{Robson}. Especially, the structure of $^{12}$C has received a lot of interest ranging from studies based on Antisymmetric Molecular Dynamics \cite{AMD}, Fermion Molecular Dynamics \cite{FMD}, BEC-like cluster model \cite{BEC,Schuck}, ab initio no-core shell model \cite{NCSM1,NCSM2,NCSM4}, lattice EFT \cite{EFT1,EFT2}, no-core symplectic model \cite{LSU}, the Skyrme model \cite{skyrme}, energy density functionals \cite{EDF}, state-of-the-art Monte-Carlo Shell Model calculations \cite{otsuka} and the Algebraic Cluster Model (ACM) \cite{ACM1,ACM2,PPNP}. Recent reviews on $\alpha$ clustering in nuclei can be found in Refs.~\cite{FreerFynbo,Freer}.

An interesting question concerns to what extent the cluster structure survives under the addition of extra nucleons. The aim of this article is to address this question for the case of $^{12}$C and $^{13}$C. The splitting of single-particle levels in the deformed field generated by a cluster of the three $\alpha$-particles with ${\cal D}_{3h}$ triangular symmetry was discussed in the framework of the Cluster Shell Model (CSM) \cite{CSM1}.  A study of both the rotation-vibration spectra and the electromagnetic transition rates showed that the $^{13}$C can be considered as a system of three $\alpha$-particles in a triangular configuration plus an additional neutron moving in the deformed field generated by the cluster characterized by ${\cal D}'_{3h}$ symmetry \cite{bijker5}.

The main idea of this article is to make a comparison between electromagnetic properties of $^{12}$C and $^{13}$C and to establish explicit relations between Coulomb form factors and transition rates in both nuclei. These relations are used to address the question to what extent the $\alpha$-cluster structure as observed in $^{12}$C is still present in the neighboring nucleus $^{13}$C. Hereto we present a simultaneous study of longitudinal (Coulomb) form factors and $B(EL)$ values in $^{12}$C and $^{13}$C in the framework of an algebraic approach to clustering (ACM and CSM) which emphasizes the symmetry structure of the cluster configurations. We derive explicit expressions between longitudinal form factors and transition probabilities in $^{13}$C which are similar to the Alaga rules in the collective geometric model. Moreover, it will be shown that since the Coulomb form factors are dominated by the collective (or cluster) part, the $q$ dependence of the longitudinal form factors in $^{12}$C and $^{13}$C is expected to be very similar.

\section{Odd-cluster nucleus: $^{13}$C}

The cluster shell model was introduced \cite{CSM1,CSM2,bijker5,PPNP} to describe nuclei composed of $k$ $\alpha$-particles plus additional nucleons, denoted as $k \alpha + x$ nuclei. The CSM combines cluster and single-particle degrees of freedom, and is very similar in spirit as the Nilsson model \cite{Nilsson}, but in the CSM the odd nucleon moves in the deformed field generated by the (collective) cluster degrees of freedom. 
For the case of $^{13}$C the total Hamiltonian is given by
\ba
H=H_{\rm int}+H_{\rm rot} ~,
\ea
where $H_{\rm int}$ is the intrinsic single-particle CSM Hamiltonian \cite{CSM1,bijker5}
\ba
H_{\rm int} = T + V(\vec{r}) + V_{\rm so}(\vec{r}) + \frac{1}{2}(1+\tau_3) V_{\rm C}(\vec{r}) ~,
\label{hcsm}
\ea
{\it i.e.} the sum of the kinetic energy, a central potential obtained by convoluting the density 
\ba
\rho(\vec{r}) = \left( \frac{\alpha}{\pi}\right)^{3/2} 
\sum_{i=1}^{3}\exp \left[ -\alpha \left( \vec{r}-\vec{r}_{i}\right)^{2}\right] ~,  
\label{rhor}
\ea
with the interaction between the $\alpha$-particle and the nucleon, a spin-orbit interaction and, for an odd proton, a Coulomb potential. Here $\vec{r}_i=(r_i,\theta_i,\phi_i)$ denote the coordinates of the $\alpha$-particles with respect to the center-of-mass. For the case of a triangular configuration the relative distance of the three $\alpha$-particles to the center-of-mass is the same $r_i=\beta$, and the coordinates are given by $(\beta,\frac{\pi}{2},0)$, $(\beta,\frac{\pi}{2},\frac{2\pi}{3})$ and $(\beta,\frac{\pi}{2},\frac{4\pi}{3})$.

The CSM makes use of a symmetry-adapted basis for ${\cal D}_{3h}$ symmetry instead of a spherical basis \cite{bijker5,adrian}. For the case of triangular symmetry the eigenstates of the CSM Hamiltonian of Eq.~(\ref{hcsm}) can be classified according to the doubly degenerate spinor representations of the double point group ${\cal D}'_{3h} $\cite{Herzberg3}: $\Omega=E_{1/2}$ with $\gamma=\pm 1/2$, $\Omega=E_{5/2}$ with $\gamma=\pm 5/2$ or $\Omega=E_{3/2}$ with $\gamma=\pm 3/2$  or, in the notation of Ref.~\cite{bijker5}, as $\Omega=E_{1/2}^{(+)}$, $E_{1/2}^{(-)}$ and $E_{3/2}$, respectively. The label $\gamma$ was introduced to distinguish between the two components of each one of the doublets \cite{adrian}. 
The Hamiltonian of the CSM is solved in the intrinsic, or body-fixed, system. 
The single-particle energies and the intrinsic wave functions are obtained by diagonalizing 
the $H$ in the harmonic oscillator basis $| nljm \rangle$ 
\ba
\phi_{\Omega\gamma} = \sum_{nljm} C^{\Omega\gamma}_{nljm} \left| nljm \right> ~.
\ea
The rotational states are labeled by the angular momentum $I$, parity $P$ and its projection $K$ on the symmetry axis, $| I^P K \rangle$. The allowed values of $K^P$ for each one of the spinor representations are given by \cite{bijker5,adrian}
\ba
\begin{array}{lcl}
\Omega=E_{1/2}, \;\gamma=\pm\frac{1}{2} &:& K^P = \pm\frac{1}{2}^+, \mp\frac{5}{2}^-, \pm\frac{7}{2}^-, \ldots \\ 
\Omega=E_{5/2}, \;\gamma=\pm\frac{5}{2} &:& K^P = \mp\frac{1}{2}^-, \pm\frac{5}{2}^+, \mp\frac{7}{2}^+, \ldots \\ 
\Omega=E_{3/2}, \;\gamma=\pm\frac{3}{2} &:& K^P = \pm\frac{3}{2}^+, \mp\frac{3}{2}^-, \mp\frac{9}{2}^+, \ldots 
\end{array}
\ea
with $I=|K|$, $|K|+1$, $\ldots$. The ground state band in $^{13}$C has $E_{5/2}$ symmetry under ${\cal D}'_{3h}$ and consists of several rotational bands labeled by $|K|^P=\frac{1}{2}^-$, $\frac{5}{2}^+$, $\frac{7}{2}^+$, $\ldots$ \cite{bijker5}. The complete wave function is given by \cite{adrian}
\ba
\left| \Omega\gamma;I^PMK \right> = \sqrt{\frac{2I+1}{16\pi^2}} \psi_v 
\left( 1 + {\cal S}_i^{-1} {\cal S}_e \right) \phi_{\Omega\gamma} D_{MK}^I(\omega) ~,
\ea
where $\psi_v$ is the vibrational wave function, $\phi_{\Omega\gamma}$ the intrinsic wave function and $D_{MK}^I(\omega)$ the rotational wave function. The wave function is invariant under the transformation ${\cal S}_i^{-1} {\cal S}_e$ where the operator ${\cal S}$ is the product of a rotation about $\pi$ followed by a parity transformation \cite{adrian}. The operator ${\cal S}_i$ acts on the intrinsic wave function and ${\cal S}_e$ on the rotational wave function.

\section{Longitudinal form factors}

For a system consisting of a cluster of three $\alpha$-particles in a triangular 
geometry and a single nucleon 
the charge distribution is taken be the sum of a Gaussian-like distribution for the 
three $\alpha$-particles and a point-like distribution for the extra nucleon
\ba
\rho_{\rm ch}(\vec{r}) &=& \rho_{\rm ch}^{\rm c}(\vec{r}) + \rho_{\rm ch}^{\rm sp}(\vec{r}) ~,
\ea
with
\ba
\rho_{\rm ch}^{\rm c}(\vec{r}) &=& \frac{(Ze)_{\rm c}}{3} \left( \frac{\alpha }{\pi} \right)^{3/2} 
\sum_{i=1}^{3} \mbox{e}^{-\alpha( \vec{r}-\vec{r}_{i})^{2}} ~,
\nonumber\\
\rho_{\rm ch}^{\rm sp}(\vec{r}) &=& \tilde{e} \, \delta(\vec{r}-\vec{r}_{\rm sp}) ~.
\ea
Here $(Ze)_{\rm c}$ is the electric charge of the $3\alpha$ core nucleus and $\tilde{e}$ denotes 
the effective charge of the extra nucleon. The longitudinal or Coulomb form factor 
$F_{C\lambda}$ is related to the matrix element of the Fourier transform of the 
charge distribution
\ba
A = \frac{\left< \Omega\gamma;I^PKM \left| 
\int \rho_{\rm ch}(\vec{r}) \mbox{e}^{i \vec{q} \cdot \vec{r}} d\vec{r} \, \right| \Omega'\gamma';I'^{P'}K'M' \right>}{(Ze)_{\rm odd}} ~,
\ea
summed over final and averaged over initial states
\ba
\frac{1}{2I'+1} \sum_{MM'} \left| A \right|^2 =
\sum_\lambda \left| F_{C\lambda}(q;\Omega'\gamma',I'K' \rightarrow \Omega\gamma,IK) \right|^2 
\ea 
with
\ba
&& F_{C\lambda}(q;\Omega'\gamma',I'K' \rightarrow \Omega\gamma,IK) 
\nonumber\\
&& \quad = \frac{\sqrt{4\pi}}{(Ze)_{\rm odd}} \Big[ \left< I',K',\lambda,K-K' | I,K \right>  
\nonumber\\
&& \qquad \quad \left\{ \delta_{\Omega\Omega'} \delta_{\gamma\gamma'} G^{\rm c}_{vv'}(q)
+ \delta_{vv'} G^{\rm sp}_{\Omega\gamma,\Omega'\gamma'}(q) \right\} 
\nonumber\\
&& \qquad \quad + P(-1)^{I+K} \left< I',K',\lambda,-K-K' | I,-K \right> 
\nonumber\\
&& \qquad \quad \delta_{vv'} H^{\rm sp}_{\Omega\gamma,\Omega'\gamma'}(q) \Big] ~.
\ea
$(Ze)_{\rm odd}$ denotes the electric charge of the odd nucleus.

For the vibrationally elastic case with $v=v'=0$ the collective part is given by 
\ba
G^{\rm c}_{00}(q;\lambda,K-K') &=& \int \rho_{\rm ch}^{\rm c}(\vec{r}) j_\lambda(qr) Y_{\lambda,K-K'}(\hat r) d\vec{r} 
\nonumber\\
&=& (Ze)_{\rm c} \, j_\lambda(q\beta) \, \mbox{e}^{-q^2/4\alpha} 
\nonumber\\
&& Y_{\lambda,K-K'}(\tfrac{\pi}{2},0) \, \delta_{K-K',3\kappa} ~.
\label{gcoll}
\ea
The single-particle part is obtained in the CSM by
\ba
G^{\rm sp}_{\Omega\gamma,\Omega'\gamma'}(q) &=& 
\sum_{nljm} \sum_{n'l'j'm'}  C^{\Omega\gamma}_{nljm} C^{\Omega'\gamma'}_{n'l'j'm'} \nonumber\\
&&\left< nljm \left| \hat M_{\lambda,K-K'}(\vec{r}) \right| n'l'j'm' \right> ~,
\nonumber\\
H^{\rm sp}_{\Omega\gamma,\Omega'\gamma'}(q) &=& 
\sum_{nljm} \sum_{n'l'j'm'}  C^{\Omega\gamma}_{nljm} C^{\Omega'\gamma'}_{n'l'j'm'} 
(-1)^{n+j+m} 
\nonumber\\
&& \left< nlj,-m \left| \hat M_{\lambda,-K-K'}(\vec{r}) \right| n'l'j'm' \right> ~,
\nonumber\\
\hat M_{\lambda\mu}(\vec{r}) &=& \tilde{e} \, j_\lambda(qr) \, Y_{\lambda\mu}(\hat r) ~.
\label{ghsp}
\ea

For longitudinal form factors which are diagonal in the intrinsic states, the contribution of the term $H^{\rm sp}$ vanishes identically. 
As a consequence, ratios of these diagonal longitudinal form factors do not depend 
on the momentum transfer $q$, and are given simply by ratios of Clebsch-Gordan coefficients.
These relations are valid for both the collective and the single-particle part, and are similar, but not identical, to the well-known Alaga rules in the collective geometric model \cite{alaga}. For excitations from the ground state of $^{13}$C with $\Omega= E_{5/2}$, $\gamma =-\frac{5}{2}$ and $I'=K'=\frac{1}{2}$ to final states with $I_1=\lambda-\frac{1}{2}$, $K_1$ relative to the final state with $I_2=\lambda+\frac{1}{2}$, $K_2$ the ratio is given by 
\ba
\frac{\left| F_{C\lambda}(q;\Omega\gamma,\frac{1}{2},\frac{1}{2} \rightarrow \Omega\gamma,\lambda-\frac{1}{2},K_1) \right|^2} 
{\left| F_{C\lambda}(q;\Omega\gamma,\frac{1}{2},\frac{1}{2} \rightarrow \Omega\gamma,\lambda+\frac{1}{2},K_2) \right|^2} 
= \frac{\lambda-K_1+\frac{1}{2}}{\lambda+K_2+\frac{1}{2}} ~.
\label{FCrel}
\ea
The explicit results are shown in the last column of Table~\ref{alagatab}.

\section{Even-cluster nucleus: $^{12}$C}

The even-cluster nucleus $^{12}$C was described successfully in the Algebraic Cluster Model (ACM) as a triangular configuration of three $\alpha$-particles \cite{ACM2}. In particular, the $L^P=5^-$ was predicted more than a decade before its measurement \cite{marin}. The relevant point group symmetry is ${\cal D}_{3h}$. The ground state band in $^{12}$C has $A'_1 \supset A_1$ symmetry under ${\cal D}_{3h} \supset {\cal D}_3$ and consists of several rotational bands labeled by $K^P=0^+$, $3^-$, $6^+$, $\ldots$, where $K$ is the projection of the angular momentum $L$ on the symmetry axis. The angular momentum content is given by $L^P=0^+$, $2^+$, $4^+$, $\ldots$ for $K^P=0^+$ and $L^P=K^P$, $(K+1)^P$, $\ldots$ for $K \neq 0$. The longitudinal (or Coulomb) form factor $F_{C\lambda}$ for vibrationally elastic excitations from the ground state with $A'_1 \supset A_1$ symmetry is given by \cite{ACM2}
\ba
&& \left| F_{C\lambda}(q;0^+,0 \rightarrow L^P,K) \right|^2
\nonumber\\
&& \qquad \qquad = \delta_{\lambda L}  \frac{4\pi}{(Ze)^2_{\rm c}} \frac{2\left| G^{\rm c}_{00}(q;\lambda,K) \right|^2}{1+\delta_{K,0}}
\nonumber\\
&& \qquad \qquad = \delta_{\lambda L} \, c_{LK}^2 \left| j_L(q\beta) \, e^{-q^2/4\alpha} \right|^2 ~,
\label{ff12c}
\ea
where the coefficients $c_{LK}^2$ can be obtained from Eq.~(\ref{gcoll}) as
\ba
c_{LK}^2 = 4\pi \left| Y_{LK}(\tfrac{\pi}{2},0) \right|^2
\frac{2\delta_{K,3\kappa}}{1+\delta_{K,0}} ~.
\ea
Specific values are $C_{00}^2=1$, $C_{20}^2=\frac{5}{4}$, $C_{33}^2=\frac{35}{8}$, $C_{40}^2=\frac{81}{64}$, $C_{53}^2=\frac{385}{128}$, $C_{60}^2=\frac{325}{256}$ and $C_{66}^2=\frac{3003}{512}$.

The transition probabilities are obtained in the long wavelength limit as
\ba
B(E\lambda;0^+,0 \rightarrow L^P,K) = \delta_{\lambda L} \,\frac{(Ze)_{\rm c}^2 \, \beta^{2\lambda} \, c_{LK}^2}{4\pi} ~.
\ea
The classical result of multipole radiation for a system of three $\alpha$ particles located at the vertices of an equilateral triangle given in Eq.~(2.21) of Ref.~\cite{ACM2} corresponds quantummechanically to a sum over all states that can be excited from the ground state. The explicit connection is a result of the equality
\ba
\sum_{K=-L}^{L} \delta_{K,3\kappa} \, c_{LK}^2 
= \frac{(2L+1)\left[1+2P_L(-\frac{1}{2})\right]}{3} ~.
\ea

\section{Comparison between $^{12}$C and $^{13}$C}

In this section, we discuss the Coulomb form factors and the electromagnetic transition probabilities for $^{12}$C and $^{13}$C. 

\begin{table}
\centering
\caption[]{Relation between the collective contribution to the Coulomb form factors 
for excitations from the ground state in $^{12}$C with $L^P=0^+$, $K=0$ and the ground state
in $^{13}$C with $I^P=\frac{1}{2}^-$, $K=\frac{1}{2}$.}
\label{alagatab}
\vspace{10pt}
\begin{tabular}{crccr}
\hline
\noalign{\smallskip}
\multicolumn{2}{c}{$^{12}$C} && \multicolumn{2}{c}{$^{13}$C} \\
$L^P,|K|$ & $| F_{C\lambda}/\frac{\sqrt{4\pi}}{(Ze)_{\rm c}} |^2$ && 
$I^P,K$ & $| F_{C\lambda}/\frac{\sqrt{4\pi}}{(Ze)_{\rm odd}} |^2$ \\
\noalign{\smallskip}
\hline
\noalign{\smallskip}
$0^+,0$ & $\left| G^{\rm c}_{00}(q;0,0) \right|^2$ &&  
$\frac{1}{2}^-,\frac{1}{2}$ & $\left| G^{\rm c}_{00}(q;0,0) \right|^2$ \\
\noalign{\smallskip}
$2^+,0$ & $\left| G^{\rm c}_{00}(q;2,0) \right|^2$ &&  
$\frac{3}{2}^-,\frac{1}{2}$ & $\frac{2}{5} \left| G^{\rm c}_{00}(q;2,0) \right|^2$ \\
&&& $\frac{5}{2}^-,\frac{1}{2}$ & 
$\frac{3}{5} \left| G^{\rm c}_{00}(q;2,0) \right|^2$ \\
\noalign{\smallskip}
$4^+,0$ & $\left| G^{\rm c}_{00}(q;4,0) \right|^2$ && 
$\frac{7}{2}^-,\frac{1}{2}$ & 
$\frac{4}{9} \left| G^{\rm c}_{00}(q;4,0) \right|^2$ \\
&&& $\frac{9}{2}^-,\frac{1}{2}$ & 
$\frac{5}{9} \left| G^{\rm c}_{00}(q;4,0) \right|^2$ \\
\noalign{\smallskip}
$3^-,3$ & $2\left| G^{\rm c}_{00}(q;3,3) \right|^2$ && 
$\frac{5}{2}^+,-\frac{5}{2}$ & 
$\frac{6}{7} \left| G^{\rm c}_{00}(q;3,3) \right|^2$ \\
&&& $\frac{7}{2}^+,-\frac{5}{2}$ & 
$\frac{1}{7} \left| G^{\rm c}_{00}(q;3,3) \right|^2$ \\
&&& $\frac{7}{2}^+,\frac{7}{2}$ & 
$\left| G^{\rm c}_{00}(q;3,3) \right|^2$ \\
\noalign{\smallskip}
\hline
\end{tabular}
\end{table}

\subsection{Longitudinal form factors}

\begin{figure}
\centering
\vspace{15pt}
\setlength{\unitlength}{1pt}
\resizebox{10cm}{!}{
\begin{picture}(350,200)(25,0)
\thicklines
\put( 50, 50) {\line(1,0){30}} 
\put( 50, 80) {\line(1,0){30}} 
\put( 50,150) {\color{red} \line(1,0){30}} 
\put( 60, 50) {\vector(0,1){30}} 
\put( 70, 50) {\color{red} \vector(0,1){100}} 
\put( 82, 47) {$0^+$}
\put( 82, 77) {$2^+$}
\put( 82,147) {\color{red} $4^+$}
\put(100,110) {\color{blue} \line(1,0){30}}
\put(115, 50) {\color{blue} \vector(0,1){60}} 
\put(132,107) {\color{blue} $3^-$}
\put( 25, 20) {$|K|^P=0^+$}
\put(113, 20) {$3^-$}
\multiput(98, 50)(4,0){10}{\circle*{1}}

\put(180, 50) {\line(1,0){30}} 
\put(180, 65) {\line(1,0){30}} 
\put(180, 90) {\line(1,0){30}} 
\put(180,125) {\color{red} \line(1,0){30}} 
\put(180,170) {\color{red} \line(1,0){30}} 
\put(185, 50) {\vector(0,1){15}} 
\put(190, 50) {\vector(0,1){40}} 
\put(200, 50) {\color{red} \vector(0,1){75}} 
\put(205, 50) {\color{red} \vector(0,1){120}} 
\put(212, 47) {$\frac{1}{2}^-$}
\put(212, 62) {$\frac{3}{2}^-$}
\put(212, 87) {$\frac{5}{2}^-$}
\put(212,122) {\color{red} $\frac{7}{2}^-$}
\put(212,167) {\color{red} $\frac{9}{2}^-$}

\put(230, 90) {\color{blue} \line(1,0){30}} 
\put(230,125) {\color{blue} \line(1,0){30}} 
\put(240, 50) {\color{blue} \vector(0,1){40}} 
\put(250, 50) {\color{blue} \vector(0,1){75}} 
\put(262, 87) {\color{blue} $\frac{5}{2}^+$}
\put(262,122) {\color{blue} $\frac{7}{2}^+$}
\put(280,125) {\color{blue} \line(1,0){30}} 
\put(295, 50) {\color{blue} \vector(0,1){75}} 
\put(312,122) {\color{blue} $\frac{7}{2}^+$}

\put(193, 20) {$\frac{1}{2}^-$}
\put(243, 20) {$\frac{5}{2}^+$}
\put(293, 20) {$\frac{7}{2}^+$}
\multiput(228, 50)(4,0){22}{\circle*{1}}
\end{picture}}
\caption[]{Non-vanishing Coulomb form factors for $C2$ (black), $C3$ (red) and $C4$ (blue) vibrationally elastic excitations from the ground state of $^{12}$C (left)  and $^{13}$C (right).}
\label{Coulomb}
\end{figure}
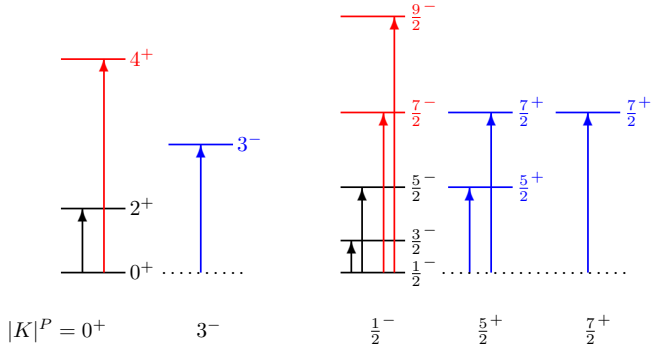

There is a close relation between the Coulomb form factors and transition probabilities in the nuclei $^{12}$C and $^{13}$C. Fig.~\ref{Coulomb} shows the non-vanishing Coulomb form factors for vibrationally elastic excitations ($v=v'=0$ and $\Omega'\gamma'=\Omega\gamma$) from the ground state in both nuclei. For a given multipole the relative ratios for Coulomb form factors to excited states in $^{13}$C are given by Eq.~(\ref{FCrel}). In $^{12}$C only a single state is excited for multipoles $C2$, $C3$ and $C4$. Even though in $^{13}$C the strength is fragmented, it is only fragmented over a limited number of excited states, two states for $C2$ and $C4$ and 3 states for $C3$. 

In Table~\ref{alagatab}, we show the relation between the Coulomb form factors in $^{12}$C and $^{13}$C. The longitudinal or Coulomb form factors are dominated by the collective part. Since the contribution of the single-particle part is of the order of a few percent only, it is neglected in the table. For a given multipole the $C\lambda$ Coulomb form factors in $^{12}$C and $^{13}$C all have the same $q$ dependence, and the summed strength for the excitation from the ground state to excited states is the same in both nuclei
\ba
&& \left| F_{C\lambda}(q;0^+,0 \rightarrow L^P,K) \right|^2 
\nonumber\\
&& \qquad = \sum_{I,K'} \left| F_{C\lambda}(q;\tfrac{1}{2}^-,\tfrac{1}{2} \rightarrow I^{P'},K') \right|_{\rm c}^2 ~,
\label{ff12C13C}
\ea
where $K'=\frac{1}{2}$ for $K=0$ and $K'=K\pm\frac{1}{2}$ for $K\neq0$. The final states in $^{12}$C have parity $P=(-1)^\lambda$ and those in $^{13}$C $P'=-(-1)^\lambda$. 

\begin{figure}[!]
\centering
\includegraphics[scale=0.5]{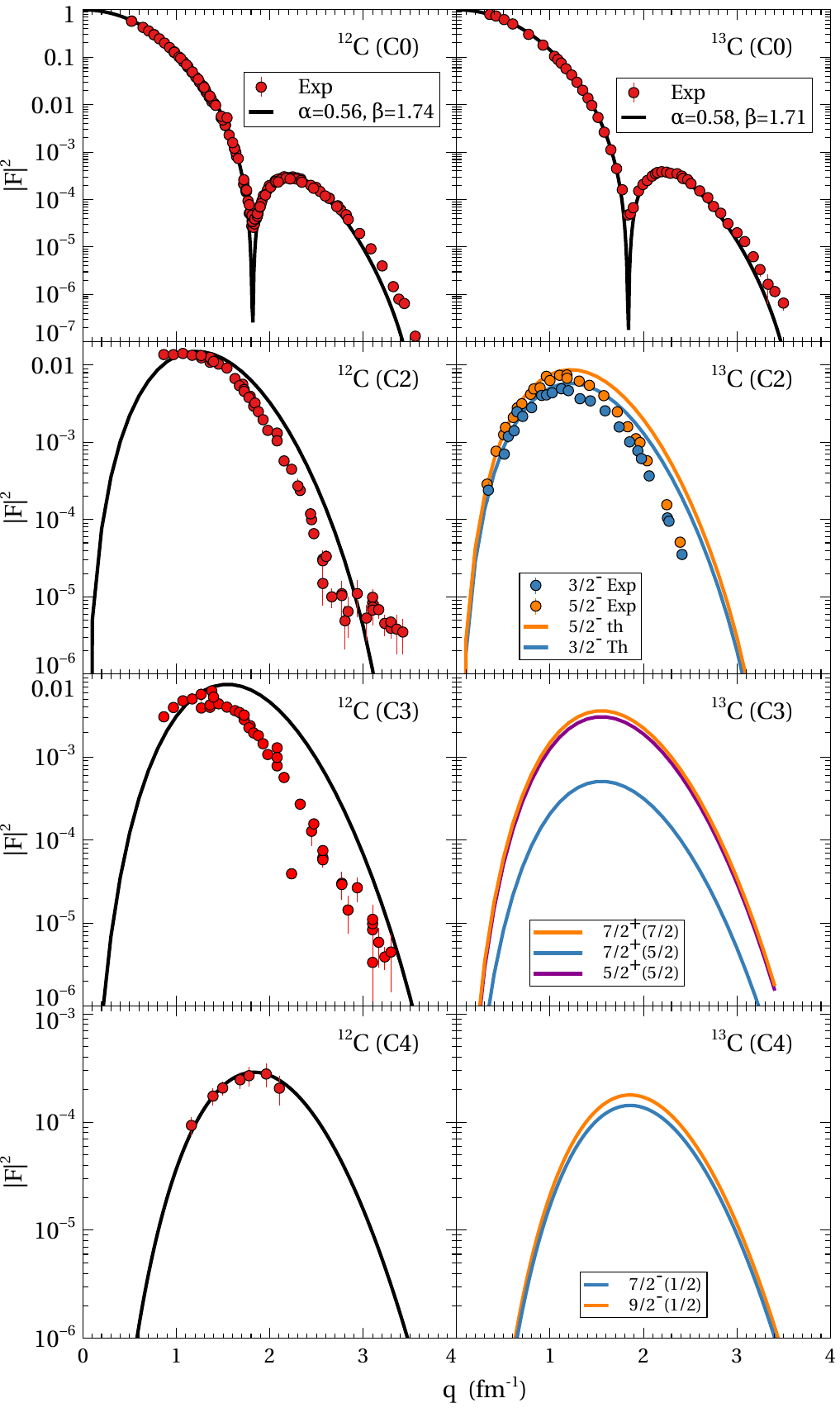}
\caption[]{\label{ffclam}Longitudinal C0, C2, C3 and C4 form factors in $^{12}$C (left) and $^{13}$C (right). The experimental data are taken from \cite{reuter,sick,crannell1,crannell2,nakada,elff,inelff}.}
\end{figure}

\begin{figure}[!]
\centering
\includegraphics[scale=0.5]{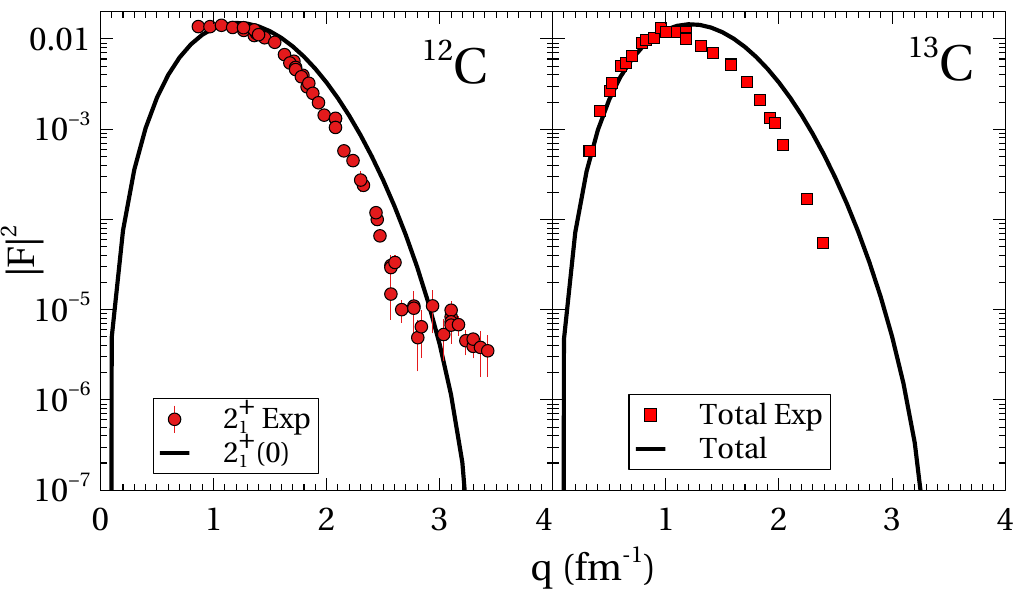}
\caption[]{\label{ffc2}C2 form factor in $^{12}$C (left) and $^{13}$C(right). For $^{13}$C the squares represent the sum of the C2 form factors to the states $\frac{5}{2}_1^-$ and $\frac{3}{2}_1^-$. The experimental data are taken from \cite{crannell1,crannell2} and \cite{inelff}.}
\end{figure}

In Fig.~\ref{ffclam} we show a comparison of the elastic form factors in $^{12}$C (left) and $^{13}$C (right). The coefficients $\alpha$ and $\beta$ are determined from the elastic form factors to be $\alpha=0.56$ 1/fm$^2$ and $\beta=1.74$ fm for $^{12}$C, and $\alpha=0.58$ 1/fm$^2$ and $\beta=1.71$ fm for $^{13}$C. The effective charge of the odd neutron is taken as the center-of-mass corrected value, $\tilde{e}=\left(-\right)^{\lambda} Ze/A^{\lambda}$ \cite{CSM2}. 
The similarity between the $C0$ and $C2$ form factors in $^{12}$C and $^{13}$C is a consequence of the cluster structure in both nuclei. Since for these states the Coulomb form factors are dominated by the collective (or cluster) part, the $q$ dependence of the longitudinal form factors in $^{12}$C and $^{13}$C is very similar.

A special case of Eq.~\ref{ff12C13C} can be seen in Fig.~\ref{ffc2} in which we compare the summed $C2$ strength in both nuclei. Whereas in $^{12}$C only the $L^P=2^+$ state is excited, in $^{13}$C the strength is fragmented among the $\frac{3}{2}_1^-$ and $\frac{5}{2}_1^-$ states. 
The experimental data in Figs.~\ref{ffclam} and \ref{ffc2} show indeed that the $q$ dependence is very similar for all three form factors, and hence also for the summed strength, in agreement 
with the CSM results in Eq.~\ref{ff12C13C}.

\subsection{Transition probabilities}

The transition probabilities $B(E\lambda)$ can be obtained from the Coulomb form factors in the long wavelength limit. Therefore, the $B(E\lambda\uparrow)$ values for excitations from the ground state of $^{12}$C and $^{13}$C satisfy the same relation as the Coulomb form factors in Eq.~(\ref{ff12C13C}). The octupole and hexadecupole transitions are related to the quadrupole transitions by the value of $\beta$
\ba
\frac{B(E3;3^{-},3 \rightarrow 0^{+},0)}{B(E2;2^{+},0 \rightarrow 0^{+},0)} 
&=& \frac{5}{2}  \beta^2  
\nonumber\\
\frac{B(E4;4^{-},0 \rightarrow 0^{+},0)}{B(E2;2^{+},0 \rightarrow 0^{+},0)} 
&=& \frac{9}{16}  \beta^4
\ea
In Table~\ref{bel0} we present a comparison between calculated $B(E\lambda)$ values and quadrupole moments and experimental data. There is good agreement between the calculations and the experimental data. The quadrupole transitions from the $\frac{5}{2}_1^-$ and $\frac{3}{2}_1^-$ states to the ground state have very similar values as is predicted in the CSM. 

\begin{table}[!]
\centering
\caption[]{$B(E\lambda)$ values and quadrupole moments 
in $^{12}$C and $^{13}$C. Experimental data are taken from \cite{DAlessio,ajzenberg}.}
\label{bel0}
\vspace{10pt}
\resizebox{\linewidth}{!}{
\begin{tabular}{lcrcl}
\hline
\noalign{\smallskip}
& $B(E\lambda)$ & Th & Exp & \\ 
\noalign{\smallskip}
\hline
\noalign{\smallskip}
$^{12}$C & $B(E2;2^{+},0 \rightarrow 0^{+},0)$ & $6.6$ & $7.63 \pm 0.19$ & $e^2\mbox{fm}^4$ \\ 
\noalign{\smallskip}
& $B(E3;3^{-},3 \rightarrow 0^{+},0)$ & $49.7$ & $103 \pm 17$ & $e^2\mbox{fm}^6$ \\ 
\noalign{\smallskip}
& $B(E4;4^{+},0 \rightarrow 0^{+},0)$ & $33.8$ & $$ & $e^2\mbox{fm}^8$ \\ 
\noalign{\smallskip}
& $Q(2^{+},0)$ & $5.2$ & $5.97 \pm 0.30$ & $e\mbox{fm}^{2}$ \\
\noalign{\smallskip}
\hline
\noalign{\smallskip}
$^{13}$C & $B(E2;\frac{3}{2}^{-},\frac{1}{2} \rightarrow \frac{1}{2}^{-},\frac{1}{2})$ & $6.1$ & $6.4 \pm 1.5$ & $e^2\mbox{fm}^4$ \\ 
\noalign{\smallskip}
& $B(E2;\frac{5}{2}^{-},\frac{1}{2} \rightarrow \frac{1}{2}^{-},\frac{1}{2} )$ & $6.1$ & $5.6 \pm 0.4$ & $e^2\mbox{fm}^4$ \\ 
\noalign{\smallskip}
& $B(E3;\frac{5}{2}^{+},-\frac{5}{2} \rightarrow \frac{1}{2}^{-},\frac{1}{2})$ & $44.7$ & $100 \pm 40$ & $e^2\mbox{fm}^6$ \\
\noalign{\smallskip}
& $B(E3;\frac{7}{2}^{+},-\frac{5}{2} \rightarrow \frac{1}{2}^{-},\frac{1}{2})$ & $5.6$ && $e^2\mbox{fm}^6$ \\
\noalign{\smallskip}
& $B(E3;\frac{7}{2}^{+},\frac{7}{2} \rightarrow \frac{1}{2}^{-},\frac{1}{2})$ & $39.1$ && $e^2\mbox{fm}^6$ \\
\noalign{\smallskip}
& $B(E4;\frac{7}{2}^{-},\frac{1}{2} \rightarrow \frac{1}{2}^{-},\frac{1}{2})$ & $29.5$ && $e^2\mbox{fm}^8$ \\
\noalign{\smallskip}
& $B(E4;\frac{9}{2}^{-},\frac{1}{2} \rightarrow \frac{1}{2}^{-},\frac{1}{2})$ & $29.5$ && $e^2\mbox{fm}^8$ \\
\noalign{\smallskip}
& $Q(\frac{5}{2}^{-},\frac{1}{2})$ & $5.0$ & & $e\mbox{fm}^{2}$ \\
\noalign{\smallskip}
& $Q(\frac{3}{2}^{-},\frac{1}{2})$ & $3.5$ & & $e\mbox{fm}^{2}$ \\
\noalign{\smallskip}
\hline
\end{tabular}}
\end{table}

It is important to note that in the present calculation all states are characterized by the same value of $\beta$, the relative distance of the $\alpha$-particles with respect to the center of mass. The value of $\beta$ was determined by the zero of the elastic Coulomb form factor, {\it i.e.} by ground state properties. It was shown by Yamada {\it et al.} that the nuclear radius for excited states is larger than that of the ground state \cite{Yamada2012}. 
In the present approach this would imply a state-dependent value of $\beta$. A slightly higher value of $\beta$ for the $L^P=3^-$ state by 10 \% would augment the $B(E3)$ value which depends on $\beta^6$ by a factor of 1.8, right up to the experimental value. Moreover, Eq.~(\ref{ff12c}) shows that a larger value of $\beta$ moves the maximum of the $C3$ form factor to a smaller $q$ value thus improving the agreement with experimental data. Since the main idea of this article was to establish explicit relations between Coulomb form factors and transition rates in $^{12}$C and $^{13}$C in order to study the extent in which the $\alpha$-cluster structure as observed in $^{12}$C is still present in the neighboring nucleus $^{13}$C, we have used a single value of $\beta$.

\section{Summary and conclusions}

We investigated to what extent the cluster structure in $^{12}$C survives under the addition of an extra neutron. Hereto we presented a simultaneous analysis of longitudinal form factors and transition probabilities in $^{12}$C and $^{13}$C in the framework of the ACM and the CSM. Since the contribution of the single-particle part is small with respect to that of the collective (or cluster) part, the $q$ dependence of the form factors is expected to be very similar in $^{12}$C and $^{13}$C which is confirmed by the available experimental data for the elastic $C0$ and inelastic $C2$ form factors. It was shown that, whereas for a given multipole only a single state in $^{12}$C is excited, in $^{13}$C the strength is fragmented, but only over a few states. Moreover, the summed strength for excitations from the ground state is the same in both nuclei. 

\begin{figure}[!]
\centering
\includegraphics[scale=0.5]{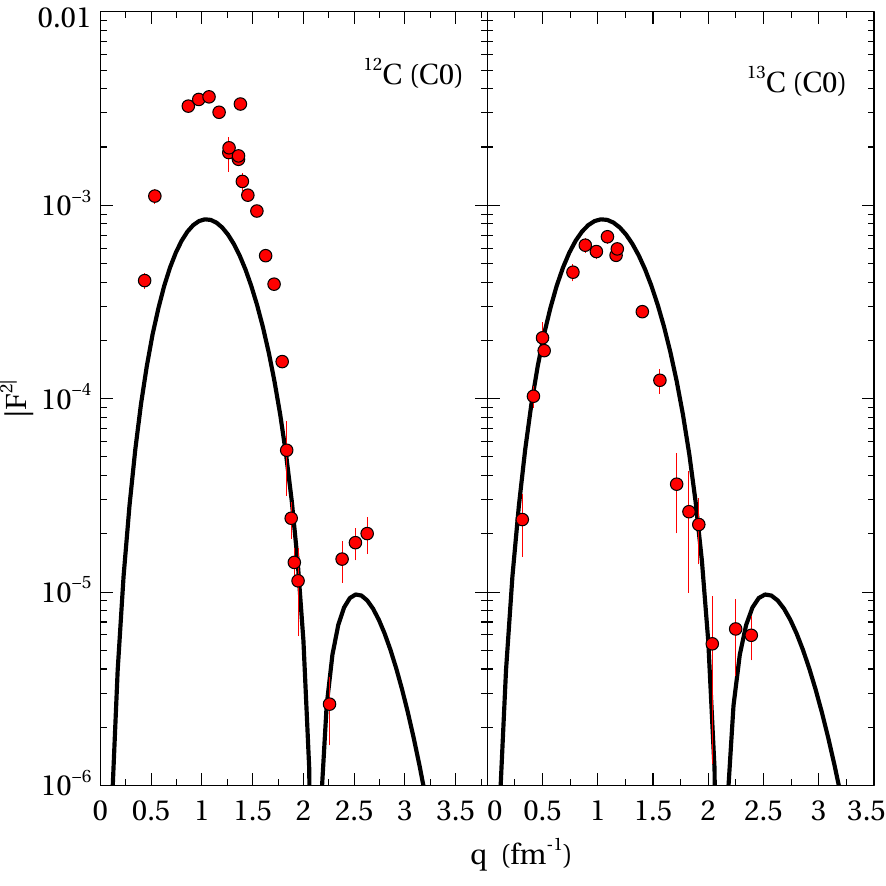}
\caption[]{C0 form factor to the $0_2^+$ Hoyle state in $^{12}$C (left) and the $\frac{1}{2}_2^-$ state at 8.86 MeV in $^{13}$C (right). The experimental data are taken from \cite{crannell1,crannell2,strehl} and \cite{inelff}. The line shows calculations taken from \cite{ACM2}, where only the core structure is considered.}
\label{ffhoyle}
\end{figure}

It is tempting to use a similar analysis to help identify the analog of the Hoyle state in $^{13}$C. In general, the search for analogs of the Hoyle state in neighboring nuclei of $^{12}$C is a topic of lot of interest, see Refs.~\cite{smith,chiba} and references therein. In Fig.~\ref{ffhoyle} we compare the longitudinal form factor of the Hoyle state with that of the $\frac{1}{2}^-_2$ state at 8.86 MeV in $^{13}$C. The $q$ dependence is very similar, although there is some discrepancy in the absolute value. The theoretical curve corresponds to a numerical calculation in the ACM \cite{ACM2}. The possible assignment of the $\frac{1}{2}^-_2$ state as the analog of the Hoyle state is in agreement with previous analyses of energy systematics \cite{bijker5,ivano}, charge radii \cite{demyanova} and spin-orbit splitting in $^{13}$C and $^{13}$N \cite{yamada}. 

In conclusion, an analysis of the Coulomb form factors and transition probabilities shows evidence for the persistence of the cluster structure of three $\alpha$-particles in a triangular configuration in the neighboring nucleus $^{13}$C.

\section*{Acknowledgements}
It is a pleasure to thank Francesco Iachello, Moshe Gai and Martin Freer for interesting discussions. This work was supported in part by research grants IN101320 and IG101423 from PAPIIT-DGAPA, UNAM, Mexico.

\bibliographystyle{elsarticle-num}
\bibliography{references}

\end{document}